\documentclass[singlecolumn]{article}
\usepackage{ae}
\usepackage{amsmath}
\usepackage{amssymb}
\usepackage{graphics}
\usepackage{graphicx}
\usepackage{epsfig}
\usepackage{dcolumn}
\usepackage{bm}

\begin{document}
\begin{center}
{\large{\bf Development of bakelite based Resistive Plate Chambers}}
\vskip 1.0 true cm

{S. Biswas$^{a}$, S. Bhattacharya$^{b}$, S. Bose$^{b}$, S.
Chattopadhyay$^{a}$, S. Saha$^{b}$, M.K. Sharan$^{b}$, Y.P.
Viyogi$^{c}$}

\vskip 1.0 true cm

{\ $^{a}$Variable Energy Cyclotron Centre, 1/AF Bidhan Nagar,
Kolkata-700
064, India}\\
{\ $^{b}$Saha Institute of Nuclear Physics, 1/AF Bidhan Nagar,
Kolkata-700
064, India}\\
{\ $^{c}$Institute of Physics, Sachivalaya Marg, Bhubaneswar,
Orissa-751 005, India}

\end{center}

\begin{center}
{\bf\noindent Abstract}\\

\end{center}

A Comparative study has been performed on Resistive Plate Chambers
made of different grades of bakelite paper laminates, produced and
commercially available in India. The chambers, operated in the
streamer mode using argon : tetrafluroethane : isobutane in 34:59:7
mixing ratio, are tested with cosmic rays for the efficiency and the
stability with cosmic rays. A particular grade of bakelite (P-120,
NEMA LI-1989 Grade XXX), used for high voltage insulation in humid
conditions, was found to give satisfactory performance with stable
efficiency of $>$ 96\% continuously for more than 110 days. A
silicone treatment of the inner surfaces of the bakelite RPC is
found to be necessary for operation of the detector.\\



{\bf\noindent 1. Introduction}\\

The Resistive Plate Chambers (RPCs), first developed by Santonico et al.
\cite{RSRC81} using bakelite and by Yu. N. Pestov et al. \cite{YNP82,WBA83}
and subsequently by others \cite{MA91},
using silicate glass are used extensively in high energy physics
experiments. The RPCs are
being considered for the following reasons a) relatively low cost
of materials used in making RPCs, b) robust fabrication procedure
and handling and c) excellent time and position resolution.
Primarily used for generating faster trigger for muon
detection \cite{GB94}, time of flight (TOF) \cite{RCRS88,AB03}
measurement, and tracking capabilities in multi layer configurations,
they are successfully used in BELLE \cite{AA02}, BaBar
\cite{BABAR95}, BESIII \cite{TBD13}, and several upcoming LHC
experiments (ATLAS,CMS etc.) \cite{ATLAS,CMS}. RPCs are used in
neutrino experiments like OPERA where its excellent time resolution
and tracking capabilities are exploited \cite{OPERA}. The
RPCs are also being explored for use in PET imaging with TOF-PET
\cite{MC07}, detection of $\gamma$-rays \cite{PC07} and neutrons
\cite{MAB03,MA03} over a large area.

The RPCs are made up of high resistive plates (e.g. glass, bakelite,
ceramics etc.) as electrodes, which help to contain the discharge
created by the passage of a charged particle or an ionizing
radiation in a gas volume, and pick-up strips are used to collect the
resulting signals. Typical time resolution for a single gap RPC is
$\sim$ 1-2 ns. By reducing the gaps between the electrodes
or by using multi-gap configuration, time resolution in such a
detector can be reduced to $<$ 100 ps \cite{PF00,ABL04}.

The RPCs are operated in two modes, viz., the proportional mode and the streamer
mode \cite{GB04}. Over the years, one of the main concerns with the use of RPCs is
their long term stability. In the proportional mode, a small amount of charge is produced
in the gas, which allows the RPC to recover in a relatively shorter
time to handle high counting rates ($\sim$ 1 KHz/cm$^{2}$). Ageing
effects caused by the accumulated charge is also relatively less in
this mode. In the streamer mode, the amount of charge produced is
considerably larger creating induced signals of larger magnitude. But,
the recovery time is larger and the irreversible damage caused by
the accumulated charge reduces the life of the RPC. However, several
remedial measures can be taken to prolong its life under streamer
mode of operation. Careful choice of materials, smoothness of the
surfaces to avoid localization of excess charges, surface treatment
to reduce the surface resistivity or providing alternate leakage path
for post-streamer recovery are adopted in the major high energy
physics experiments. Prolonged stable operation in streamer mode of the BELLE
RPCs, though made of glass, is a testimony to many serious efforts
taken for the above cause \cite{AA02}.

The glass-based RPCs are found to be more
stable mainly for low rate applications, even though some erosion
effects are found for such cases, particularly when these are
operated in the streamer mode \cite{GB04}. This has been attributed
to the corroding of the glass surface due to the large charge build
up in the streamer mode of operation. However, in the proportional
mode of operation, the detectors can be operated for longer period.
At the end of nineties, it was found that the RPCs based on
bakelites show serious ageing effects reducing the efficiency
drastically \cite{JV03}. Detailed investigations revealed that the use of
linseed oil for the surface treatment in such cases was the
main reason for this ageing effect \cite{FA03,FAN03}. Efforts were
subsequently made to look for alternatives to linseed oil treatment,
or to develop bakelite sheets which can be used without the
application of linseed oil \cite{JZ05}. It has, however, been found
that for several ongoing and future applications (e.g. CMS
detectors), bakelite based RPCs are chosen as preferred options
mainly due to cheaper cost of fabrication.

In the proposed India-based Neutrino Observatory (INO), the RPCs
have been chosen as the prime active detector for muon detection in
an Iron Calorimeter (ICAL) \cite{INO06}. As proposed presently, ICAL is a sampling
calorimeter consisting of 140 layers of magnetized iron, each of 60
mm thickness, using RPCs of 2m $\times$ 2m area as active media
sandwiched between them. A 50 Kton ICAL is expected to consist of
about 27000 RPC modules. For ICAL RPCs, main design criteria are (a)
good position resolution, (b) good timing resolution (c)
ease of fabrication in large scale with modular structure and most
importantly (d) low cost. Detailed R \& D are being performed on
glass RPCs for this application. In this article, we report a
parallel effort on building and testing of the RPC modules using the
bakelite obtained from the local industries in India. The aim of the study
is to achieve stable performance of the RPC detector for prolonged
operation.

The paper is organized as follows. In the next section, we describe
the method of assembly of the RPC modules. The section 3 contains
measurements of the bulk resistivity of the bakelite sheets used in
this application while the cosmic ray set-up used in our experiment
are discussed in section 4. The results of the study are reported in
section 5.\\

{\bf\noindent 2. Construction of the RPC modules}\\

\begin{figure}
\includegraphics[scale=1.0]{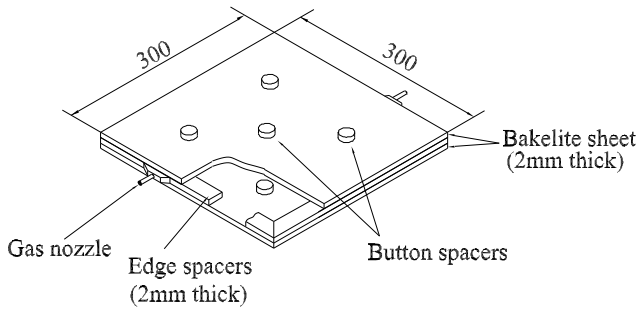}\\
\caption{\label{fig:epsart}Schematic diagram of a resistive plate
chamber.}
\end{figure}

A schematic view of the assembled RPC modules is shown in the Fig.
1. Two 300 mm $\times$ 300 mm $\times$ 2 mm bakelite sheets are used
as electrodes. The inner surfaces of the two sheets are separated by
a 2 mm gap. Uniform separation of the electrodes are ensured by
using five button spacers of 10 mm diameter and 2 mm thickness, and
edge spacers of 300 mm $\times$ 8 mm $\times$ 2 mm dimension, both
being made of polycarbonate. Two nozzles for gas inlet and
outlet, also made of polycarbonate, are placed as part of the edge
spacers. All the spacers and nozzles are glued to the bakelite
sheets using Araldite$^{\circledR}$ epoxy adhesive (grade: AY 103$^{**}$MP,
made by Huntsman A.M., (Europe)). The 2 mm thick active
gas gap of the RPC modules are leak-checked using argon and helium
sniffer probes. The edges of the bakelite sheets are sealed by
applying a layer of the epoxy adhesive to prevent permeation of
moisture.

After proper cleaning, a graphite coating is made on the outer
surfaces of bakelite sheets to distribute the applied voltage
uniformly over the entire RPC. A gap of 10 mm from the edges to the
graphite layer is maintained to avoid external sparking.
The surface resistivity varies from 500 K$\Omega$/$\Box$ to
1 M$\Omega$/$\Box$ for different samples. The
graphite coating, applied by using a spray gun, however, results in
a non-uniformity (less than 20\%) for
a particular coated surface. Two small copper foils $\sim$ 20 $\mu$m
thick are pasted by kapton tape on both the outer surfaces for the
application of high voltage. The HV connectors are soldered on these
copper strips. Equal HVs are applied on both the surfaces.

In order to collect the accumulated induced charges, pick-up strips are
placed above the graphite coated surfaces with minimum air-gap. The pick-up strips are
made of copper (20 $\mu$m thick), pasted on one side of 10 mm thick
foam. The area of each strip is 300 mm $\times$ 30 mm with a separation of
2 mm between two adjacent strips. The pick-up strips are covered
with 100 $\mu$m thick kapton foils to insulate them from the
graphite layers. The ground plane, made of
aluminium, is pasted on the other side of the foam. The signals from
different strips are sent through a ribbon cable, followed by
RG-174/U coaxial cables using proper impedance matching.

The gases used in the RPC are mixtures of Argon, Isobutane and
Tetrafluroethane (R-134a) in varying proportion. The gases are
pre-mixed, stored in a stainless steel container and sent to the
detector using stainless steel tubes. A typical flow rate of 0.4 ml
per minute resulting in $\sim$ 3 changes of gap volume per day is
maintained by the gas delivery system. A systematic analysis was
made for R-134a and Isobutane before use in the system by a Prisma
Quadstar 422 Residual Gas Analyzer. The composition was found to be
98.83\% for R-134a with 0.75\% O$_2$ and 0.41\% N$_2$ and 98.93\%
for Isobutane with 1.07\% H$_2$.\\

{\bf\noindent 3. Measurement of bulk resistivity of bakelite}\\

The bakelite sheets are phenolic resin bonded paper laminates. In
the present work, three types of bakelite sheets have been used to build as
many modules. They are (a) mechanical grade bakelite (P-1001), (b)
Superhylam grade and (c) electrical grade (P-120).

The P-1001 and P-120 grade bakelites are manufactured by Bakelite
Hylam, India and the Superhylam grade is obtained from the other manufacturer Super
Hylam, India. The sheets of P-1001 and P-120 are matt
finished whereas superhylam is glossy finished. The P-1001 has good
mechanical properties whereas the P-120 has good mechanical and
electrical properties under humid conditions prevalent in India.

The bulk resistivity of the electrode plates of the RPC is an
important parameter \cite{GB93}. The high resistivity helps in
controlling the time resolution, singles counting rate and also
prevents the discharge from spreading through the whole gas. We have
measured the bulk resistivities of the bakelite sheets via the
measurement of the leakage current.

\begin{figure}
\includegraphics[scale=0.3]{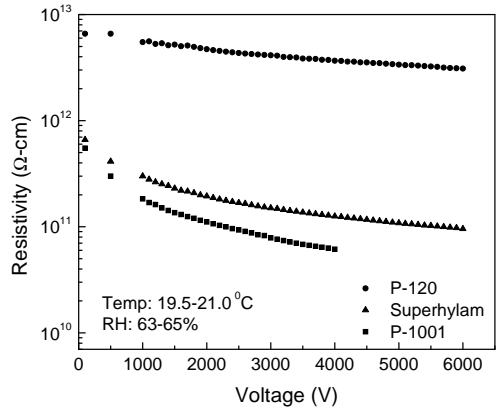}\\
\caption{\label{fig:epsart}The volume resistivity ($\rho$) as a
function of the applied voltage for three grades of bakelites.}
\end{figure}

\begin{table}
\caption{\label{tab:table1}Mechanical and electrical properties of different grades of
bakelite.}
\begin{tabular}{|c|c|c|c|c|c|c|} \hline
Trade & NEMA   &  BS-2572  & Density  &
Electrical  & Surface  & Bulk  \\
Name & LI-1989 & Grade & (g/cc) & strength & finish & resistivity  \\

  & Grade &   &   &(kV/mm)&   & ($\Omega$-cm) \\ \hline
   P-1001 & X & P1 & 1.38 & 3.5 & Matt & 6.13 $\times$ 10$^{10}$ \\ \hline
Superhylam & - & P2 & 1.72 & 9.5 & Glossy & 1.25 $\times$ 10$^{11}$ \\
\hline
 P-120 & XXX & P3 & 1.22 & 9.5 & Matt & 3.67 $\times$ 10$^{12}$   \\ \hline
\end{tabular}\\
\end{table}
This measurement is performed at the same place and the same
environment where the RPCs have been tested. The test set up is kept
in a temperature and humidity controlled room. These two parameters
have been monitored during the experiment and are nearly the same
around that time. The bulk resistivities of different grade
materials at 4 kV are tabulated in Table 1. The volume
resistivity($\rho$) vs. voltage(V) characteristics of different
grade materials are shown in Fig. 2. It is clear from the figure
that the bulk resistivity is considerably higher for the P-120 grade
bakelite. For the P-1001 grade, resistivity is much lower and it
cannot sustain high voltage above 4 kV. Therefore, the P-1001 grade
is not considered further for building up the RPC. The superhylam
grade, though having lower bulk resistivity than P-120, would stand
high voltages up to 6 kV. This is also considered for the
fabrication of RPC detector.\\

{\bf\noindent 4. Cosmic ray test setup}\\

\begin{figure}
\includegraphics[scale=0.5]{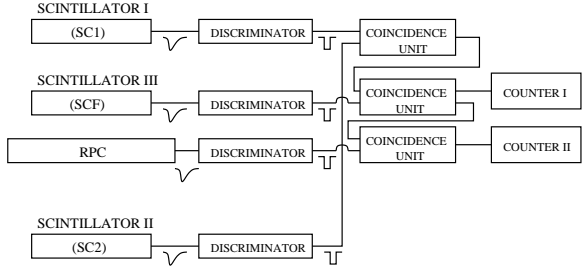}\\
\caption{\label{fig:epsart}Schematic representation of the cosmic
ray test setup.}
\end{figure}

Fig. 3 shows the schematic of the setup for testing the RPC modules
using cosmic rays. Three scintillators, two placed above the RPC plane and
one placed below are used for obtaining the trigger from the incidence of the
cosmic rays. The coincidence between scintillator I (350 mm $\times$ 250 mm
size), scintillator II (350 mm $\times$ 250 mm size) and the finger
scintillator(III) (200 mm $\times$ 40 mm size) is taken as the Master
trigger. Finally, the signal obtained from the pick-up strip of the
chamber is put in coincidence with the master trigger obtained
above. This is referred to as the coincidence trigger of the RPC.
The width of the finger scintillator is made smaller than the
total width of the two adjacent readout strips, thereby needing a
correction for dead zones in between two readout strips.

The high voltage to the RPC, are applied at the ramping
rate of 5 V/s on both the electrodes. The streamer pulses are
obtained starting from the high voltage of 5 kV across the RPC.
The high voltage is applied to some of the RPCs by using the CAEN
Mod.N470 unit and to the others using the CAEN Mod.N471A unit. The leakage
current as recorded by the high voltage system is studied. The
signals from two consecutive strips covered by the scintillator III are ORed to
form the final signal, for the next part of the pulse processing.

The Philips Scientific leading edge discriminators (Model 708) are
used for the scintillators and the RPC pulses. Various thresholds
are used on the discriminators to reduce the noise. For our final
results, a threshold of 40 mV is used on the RPC signal. We have
used a CAMAC-based data acquisition system LAMPS, developed by
Electronics Division, Bhabha Atomic Research Centre, Trombay, India.
Counts accumulated in a CAEN (Model 257) scalar over a fixed time
period are recorded at regular intervals, and saved in a periodic
log database. The temperature and the humidity are monitored at the
time of measurement.\\

{\bf\noindent 5. Results}\\

An important and obvious goal of any RPC detector development is
to study the long term stability with high efficiency. In that spirit,
the following studies are performed in the cosmic ray test bench of the RPC detectors.

The efficiency of the RPC detector, taken as the ratio between the
coincidence trigger rates of the RPC and the master trigger rates of
the 3-element plastic scintillator telescope as mentioned in sec.4, is
first studied by varying the applied HV for each detector. The rates are calculated from data
taken over 30 minutes duration for each HV setting. The temperature and
humidity during these measurement are recorded to be about 22-25$^{\circ}$C
and 63-65\% respectively. The average master trigger rate
is $\simeq$ 0.005 Hz/cm$^2$. The variation of efficiency with
applied HV is shown in the Fig. 4 and that of the singles counting rates
with the HV is shown in the Fig. 5. It is seen that for both the bakelite grades,
the efficiency has increased from 20\% to 75\% as HV is ramped up from
6.5 kV to 6.8 kV. The efficiency for the superhylam grade
gradually increases and reaches the plateau at $\sim$ 96\% from 7.5 kV, while
that of the P-120 grade reaches a maximum of $\sim$ 79\% at 7.2 kV and then
decreases steadily up to $\sim$ 35\% as the HV is increased to 9 kV. The singles
counting rates in both the cases, however, have increased more or less
exponentially with sudden jumps around 6.5-7.0 kV (see Fig. 5), i.e. near the
points where the efficiency becomes uniform (in case of superhylam) or starts to
decrease (in case of P-120). This possibly indicates the onset of a breakdown regime
that recovers in a reasonable time for the superhylam grade RPC but works the
other way for the P-120 grade bakelite RPC. It should, however, be noted that the singles
counting rate and the leakage current of the superhylam RPC are both larger than
those of the P-120 RPC, which are expected on the basis of smaller bulk
resistivity of the superhylam grade bakelite.

\begin{figure}
\includegraphics[scale=0.3]{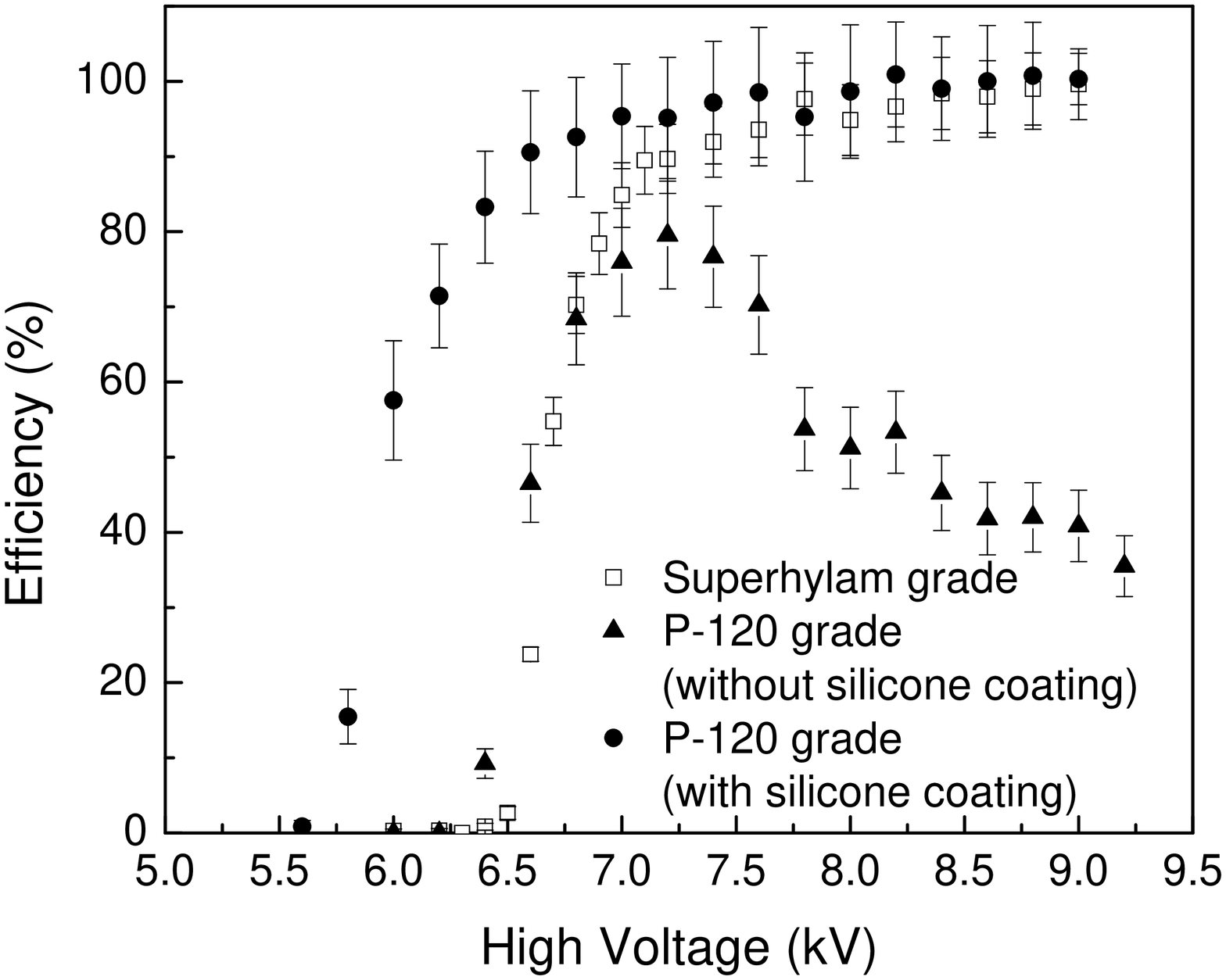}\\
\caption{\label{fig:epsart}The efficiency as a function of high
voltage for two RPCs (silicone coated \& uncoated P-120 and Superhylam grade
bakelite) obtained with a gas mixture of Argon (34\%) + Isobutane
(7\%) + R-134a (59\%). The threshold for the RPCs are set at 40
mV for P-120 and 50mV for superhylam.}
\end{figure}

\begin{figure}
\includegraphics[scale=0.3]{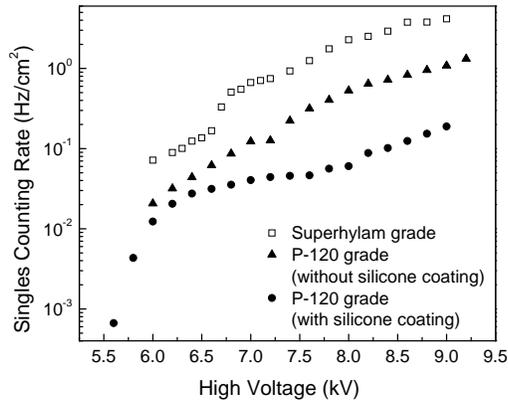}\\
\caption{\label{fig:epsart}Singles counting rate as a function of
high voltage.}
\end{figure}

In order to investigate the reason for the above phenomena, and taking
cue from the fact that superhylam surfaces are glossy finished while
the P-120 surfaces are matt finished, we have dismantled the detectors and made
surface profile scans over a 5 mm span of the surfaces using DekTak 117
Profilometer. These scans, done also for the P-1001 grade, are shown in Fig. 6.
It is clearly seen that the three surfaces have a short range variation (typically
$\sim$ 0.1 $\mu$m length scale) and a long range variation (typically
$\sim$ 1 $\mu$m length scale). The long range surface fluctuation,
which is a measure of non - uniformity, averaged over several scans are: 0.84 $\pm$ 0.12
$\mu$m (P-120), 0.49 $\pm$ 0.17 $\mu$m (superhylam) and 0.88 $\pm$ 0.09 $\mu$m (P-1001).
Thus the long range fluctuations,within the limits of
experimental uncertainties, are nearly the same. On the other hand the short range fluctuations,
a measure of surface roughness, are: 0.64 $\pm$ 0.06 $\mu$m (P-120), 0.17 $\pm$
0.02 $\mu$m (superhylam), and 0.63 $\pm$ 0.13 $\mu$m (P-1001), and thus indicate a
superior surface quality of the superhylam grade.
\begin{figure}
\includegraphics[scale=0.36]{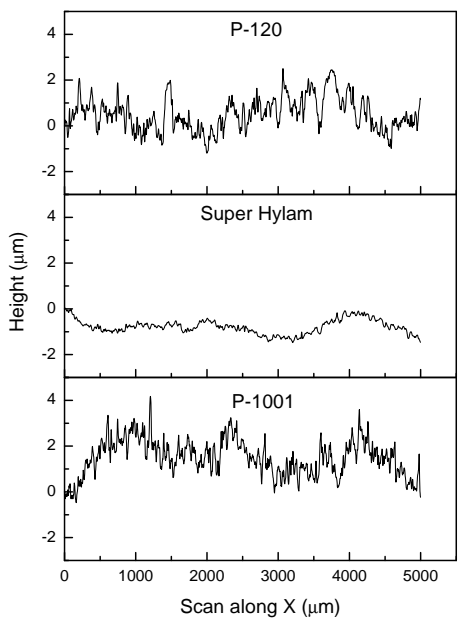}\\
\caption{\label{fig:epsart}Linear surface profile scans of the three grades of bakelite sheets.}
\end{figure}

To explore a remedial measure, we have applied a thin layer of viscous silicone
fluid (chemical formula : [R$_{2}$SiO]$_{n}$, where R = organic groups such as methyl,
ethyl, or phenyl) [coefficient of viscosity = 5500 cP, manufactured by Metroark
Limited, Kolkata, India] on the inner surfaces of the P-120 bakelite sheets.
About 1 gm of the fluid is applied over 300 mm $\times$ 300 mm area.
Based on the specific gravity (1.02 at 23$^{\circ}$C)
of the fluid, the estimated coating thickness would be $\sim$ 10 $\mu$m. This
material is chosen for the following reasons: a) very low chemical
reactivity with the gases used; b) good thermal stability over a wide
temperature range (from -100 to 250 $^{\circ}$C); c) very good electrical
insulator; d) excellent adhesion to most of the solid materials, and
e) low vapour pressure, which is essential for stable operation over
a reasonable time period. The silicone treated surfaces are kept
under infrared lamp for 24 hours to allow the viscous fluid to
fill all the micro-crevices on the surface. The reassembled
detector is tested at the same set-up. The results of
efficiency and singles count rate measurements, shown in the
Figs. 4 and 5, indicate a remarkable improvement in the performance
of the P-120 detector. The efficiency reaches from 20\%
to 75\% as the HV is increased from 5.7 kV to 6.2 kV, while the
singles count rate, as a whole has decreased by a factor of
5. This indicates quenching of micro-discharge after silicone
treatment, which is very much desirable for functioning of
the detector. The efficiency in this case reaches $>$ 95\% plateau at 7 kV.

It is worth noting that surface treatment with insulating / non-polar
liquid as a remedial measure was first demonstrated for the BaBar RPCs.
However, formation of stalagmites by polymerisation of uncured
linseed oil droplets had created conducting paths through the gap,
thereby causing irreversible damage to the bakelite plates \cite{FA03}.
The process of linseed oil treatment was later changed by
increasing the proportion of eptane as a thinner to produce a
thinner coating (10-30 $\mu$m) on the inner surface \cite{FA05}.
Our observation that silicone coating of the inner
surfaces aides the proper functioning of our P-120 bakelite
RPC detector once again confirms the importance of smooth
surface finish of the inner surfaces.

To judge the improvement in the overall performance of the RPC detector,
we have measured the leakage current through the RPC detector with and
without silicone coating and the plot of these as a function of the applied
HV as shown in the Fig. 7. Both the plots show a common feature that the
current-voltage curves have two distinctly different slopes. While the
gas gap behaves as an insulator in the lower range of applied voltage and
hence the slope over this span scales as the conductance of the polycarbonate
spacers, at higher range of voltage, the gas behaves as a conducting medium
due to the formation of the streamers. Therefore, the slope over this range scales
as the conductance of the gas gap. It is seen that the slope in the higher range
of voltage is much steeper for the RPC without silicone coating and hence it points
to the fact that some sort of uncontrolled streamers are being formed in the gas
gap causing a degradation of the efficiency. This possibly does not happen in
the RPC detector with silicone coating.

\begin{figure}
\includegraphics[scale=0.3]{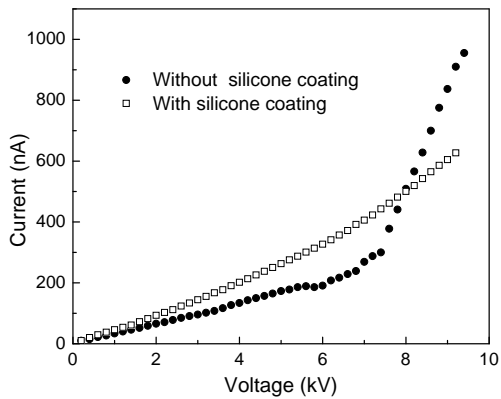}\\
\caption{\label{fig:epsart}Current as a function of the applied
voltage for RPC made by P-120 grade bakelite.}
\end{figure}

We have also examined the effect of discriminator threshold setting on the
efficiency curves of the RPC with silicone treated surfaces. These are plotted
in the Fig. 8. It is clear that the efficiency curves do not depend much on the
threshold setting from 20 mV to 80 mV, except that the efficiency plateau is
marginally higher at the lowest threshold setting of 20 mV.
\begin{figure}
\includegraphics[scale=0.3]{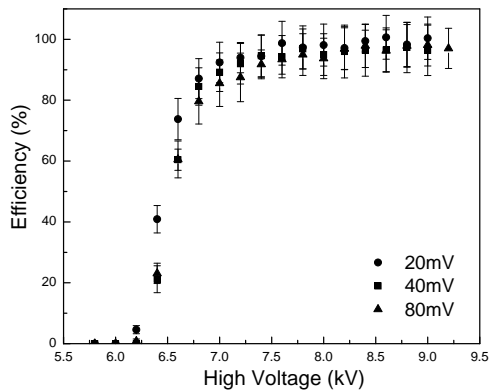}\\
\caption{\label{fig:epsart}Efficiency versus high voltage for
different thresholds for silicone coated P-120 grade bakelite RPC with silicone coating.}
\end{figure}

The effect of humidity on the efficiency curves has also been studied. This
measurement has been done at relative humidities of 58\% and 67\% of the laboratory
environment and at the same room temperature of $\sim$ 23$^{\circ}$C. These curves, plotted in the
Fig. 9, indicate no effect of humidity on the efficiency. However, the leakage currents,
measured simultaneously and plotted in the Fig. 10, are a bit larger at higher humidity.
This observation indicates that charge leakage through the exterior surfaces may be
contributing more at higher humidity.
\begin{figure}
\includegraphics[scale=0.3]{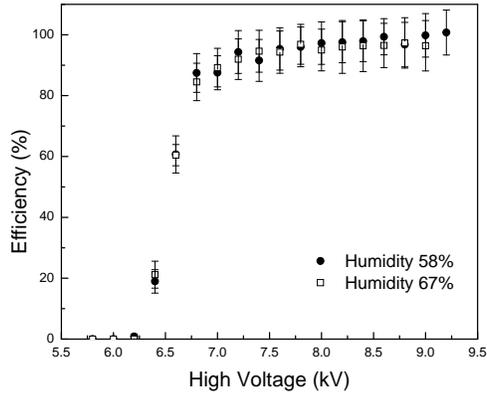}\\
\caption{\label{fig:epsart}Efficiency versus high voltage for
different humidities for silicone coated P-120 grade bakelite RPC.}
\end{figure}

\begin{figure}
\includegraphics[scale=0.3]{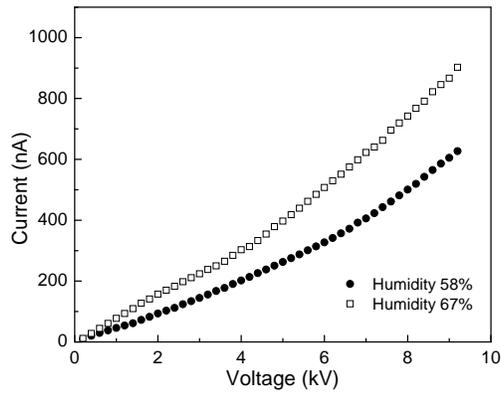}\\
\caption{\label{fig:epsart}Current versus high voltage for different
humidities for silicone coated P-120 grade bakelite RPC.}
\end{figure}

The long term stability of the bakelite RPCs has been studied using the same cosmic ray test
set-up. The coincidence trigger counts of the RPCs and the master trigger counts, accumulated
over every 2 hours, have been recorded continuously for more than 6 months. The room temperature
and the relative humidity have been controlled to keep them less
than 24$^{\circ}$C and 80\%, respectively.
When the humidity was larger, particularly during the monsoon season, the test set-up was shut
down till it came down to 80\% and below. The singles count rates of the RPCs have also been
recorded simultaneously. The Figs. 11 and 12 depict the variation of efficiency and
singles count rates over the above mentioned period for both the grades of RPCs.
The superhylam grade RPC has worked with an efficiency of $>$ 95\% which remained steady for
25 days, but beyond that, it deteriorated gradually to $\sim$ 86\%
efficiency within next 13 days. The singles count rate, however, has increased
from day one from 1 Hz/cm$^2$ to 10 Hz/cm$^2$ within 10 days, and then it
increased slowly over the next 28 days. After that period, the singles
count rate shot up to $>$ 30 Hz/cm$^2$. The leakage current gradually
increased from 3-4 $\mu$A to $>$ 10 $\mu$A within that period. This RPC
was discontinued after 38 days and the silicone coated P-120 grade RPC was then mounted. The
efficiency measured was $\sim$ 96\% and above and has remained steady
for more than 110 days. The singles count rate also has remained steady
around 0.1 Hz/cm$^2$. The leakage current was found to be marginally
dependent on temperature and humidity, though it has remained steady
at $\sim$ 400 nA during the operation.

\begin{figure}
\includegraphics[scale=0.3]{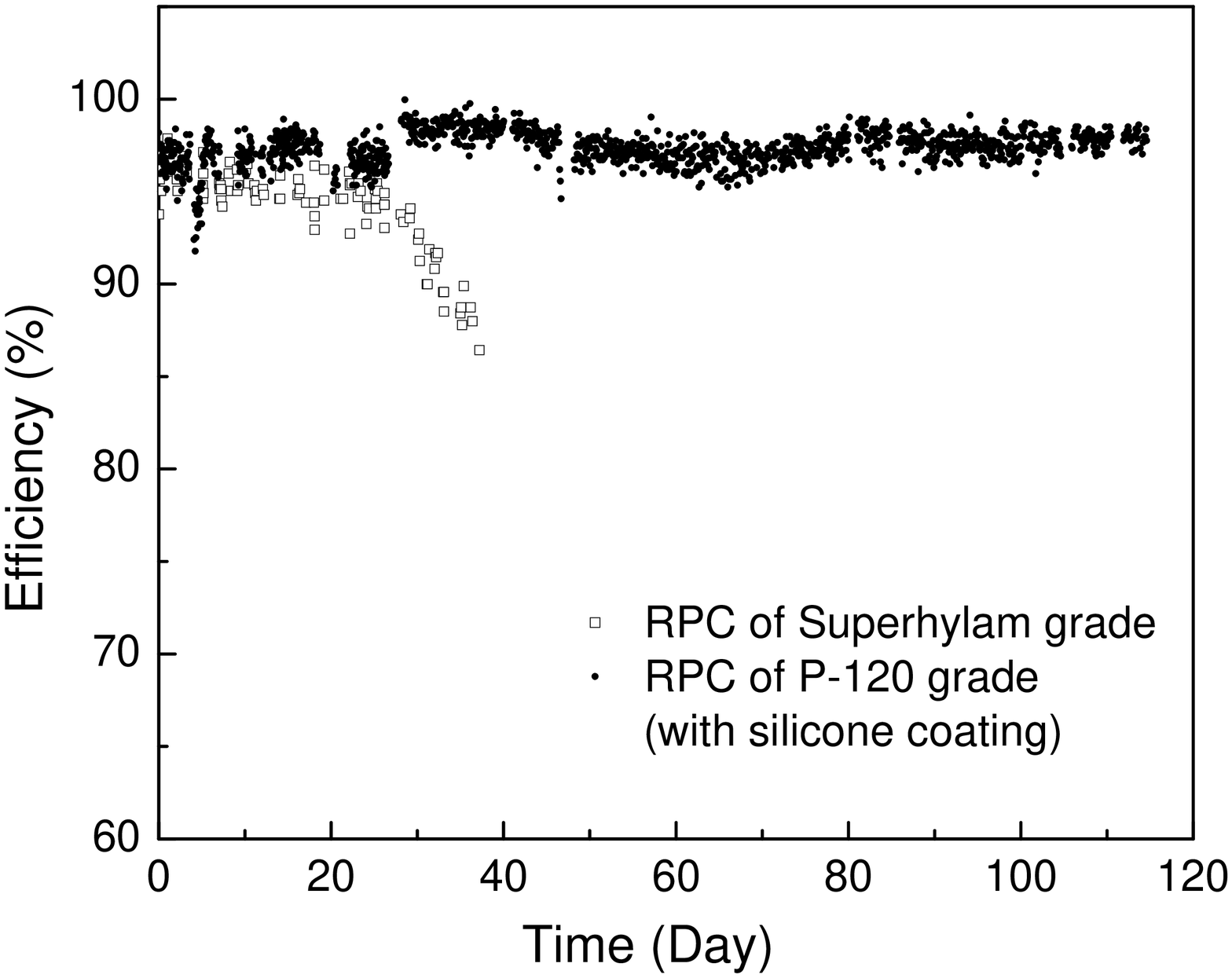}\\
\caption{\label{fig:epsart}Efficiency as a function of time for two
RPC prototypes.}
\end{figure}

\begin{figure}
\includegraphics[scale=0.3]{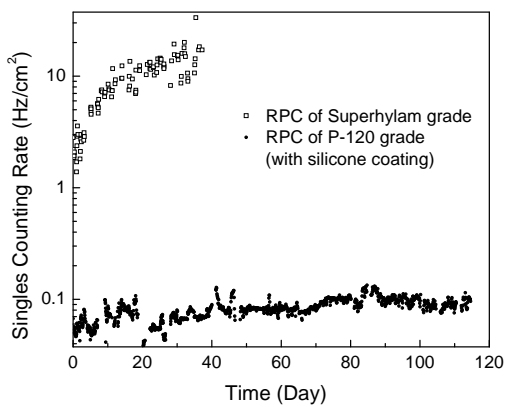}\\
\caption{\label{fig:epsart}The single counting rate as a function of
time for the two RPC prototypes.}
\end{figure}

The superhylam grade RPC has also been tested again after a gap of a
few months. It has shown the same higher leakage current ($>$ 10
$\mu$A) and lower efficiency ($\sim$ 86\%) indicating that some
intrinsic breakdown of the bulk material may have taken place.\\

{\bf\noindent 6. Conclusion}\\

We have made a comparative study of bakelite RPCs made from two
different grades of bakelite. The RPC, made of superhylam grade
bakelite with melamine coated glossy finished surface is found to
have a shorter life. On the other hand, the RPC made from P-120
grade bakelite with matt finished surfaces, which are coated with a
thin layer of viscous silicone fluid, are found to work steadily for
more than 110 days showing a constant efficiency of $>$ 96\% without
any degradation. The detector is found to be less immune to
variation in humidity which makes it a viable alternative to
semiconductive glass based RPC.\\

{\bf\noindent 7. Acknowledgement}\\

We are thankful to Prof. Naba Kumar Mandal of TIFR, India and Prof. Kazuo Abe of KEK, Japan
for their encouragement and many useful suggestions in course of this work.
We are also grateful to Dr. C. Bhattacharya, Mr. G.S.N. Murthy,
Mr. M.R. Dutta Majumdar, Mr. S.K. Thakur and Mr. S.K. Bose of VECC
for their help in the work.
We acknowledge the service rendered by Mr. Avijit Das of SINP for surface
profile scans of the bakelite sheets used by us. We would like to thank the SINP
workshop for making the components of the detectors, and Mr. Ganesh Das of VECC
for meticulously fabricating the detectors. Finally we acknowledge the
help received from the scientific staff of Electronics Workshop
Facility of SINP for building the gas flow control and delivery
system of the gas mixing unit used in this study.

\end{document}